# ACROSS-HORIZON PROPAGATION AND INFINITE TIME-DILATION: AN INDEPENDENT APPROACH TO HAWKING RADIATION

GEORGE TSOUPROS
*ATHENS - GREECE*

ABSTRACT. The conformal scalar propagator as an explicit function of the Schwarzschild black-hole space-time has been established in the preceding three projects. The present project examines the precise relation which that propagator has to the amplitude the Schwarzschild black hole emit a scalar particle in a particular mode of positive energy. It is thereby established that the effect of infinite time-dilation on the event horizon results in both, the thermal radiation in the background of a static exterior space-time geometry and the detraction from thermality if energy-conservation is properly imposed as a constraint on scalar propagation. The derivation of the latter signifies an equivalent approach to the result of Parikh and Wilczek from the semi-classical perspective of the distant observer. From that perspective for that matter, Hawking radiation emerges in both contexts as the exclusive consequence of infinite time-dilation on the event horizon. These results are consistent with Hawking radiation as the exclusive consequence of the causal structure of the Schwarzschild black-hole space-time both, in a static exterior geometry and in such dynamical exterior geometry as energy-conservation signifies. The extension of the thermal case to other spin-fields and black-hole geometries is discussed.

## I. INTRODUCTION

The intuitive approach to Hawking radiation as the result of particle-antiparticle creation in the vicinity of the event horizon receives concrete physical meaning through the Feynman propagator [2]. For a Schwarzschild black hole of mass $M$ in Planck units ($G = c = \hbar = k = 1$) this approach yields the same thermal spectrum which corresponds to Hawking temperature:

$$T_H = \frac{1}{8\pi M}$$

in the rigorous frequency-mixing approach [1]. This result has been derived in several independent treatments of Hawking radiation as quantum propagation across the event horizon (see for example [4], [5], [6], [7]). In all such approaches, Hawking radiation emerges in the context of semi-classical gravity in which all matter fields are quantised on a classical background space-time geometry.

In a remarkable development, a detraction from the thermal spectrum was established in [8] and [9] by properly imposing energy-conservation as a constraint on scalar

---

present e-mail address: landaughost@hotmail.com.

propagation. In such a treatment, Hawking radiation also emerges in the context of semi-classical gravity, albeit in a background space-time geometry which is dynamical in the exterior vicinity of the event horizon. The work in [10] transcended the essentially quantum-mechanical treatment of Hawking radiation in [8] and [9] by reproducing the stated detraction from the rigorous perspective of quantum field theory. The fact remains, nevertheless, that as long as $M$ is much bigger than the Planck mass, the rate of change of the metric $\sim M^{-3}$ is very slow compared to the typical frequency of the radiation $\sim M^{-1}$. For that matter, the assumption underlying the thermal spectrum and Hawking temperature to the effect that the background space-time geometry for $r \geq 2M$ is static remains a valid approximation [3].

The ensuing analysis uses the formalism in [2] in order to treat Hawking radiation as propagation across the event horizon in both contexts, that of the static exterior space-time and that of the dynamical space-time which energy-conservation entails. In either context, a central aspect of Hawking radiation is the fact that each particle emitted by the black hole is in an increasingly semi-classical state of very high energy when traced back toward the event horizon. Such semi-classicality emerges in two distinct perspectives and accordingly receives two equivalent interpretations. From the perspective of spatial infinity, the typical wavelength of the radiation in the vicinity of the event horizon is of the order of the size of the horizon itself and the corresponding uncertainties in the probability distributions tend to zero as a consequence of infinite time-dilation. Equivalently, the wavelength of the outgoing wave as measured by local observers situated close to the event horizon in decreasing order of the Schwarzschild radial coordinate is increasingly blue-shifted. In the vicinity of the event horizon, the radial wavenumber approaches infinity with the implication of an effectively point particle in a semi-classical state of very high energy [8].

It is in this essential respect that the ensuing analysis differs from that in [2]. The latter advances on the general expression:

$$K(x, x') = -\frac{i}{4\pi^2} \frac{1}{s(x, x') + i\epsilon}$$

for the massless conformal scalar propagator which does not involve the event horizon. For that matter, the effect of infinite time-dilation which is central to the perspective of the distant observer is not manifest. On the contrary, the present analysis advances on the propagator for a massless conformal scalar field in the Hartle-Hawking state respectively established in [11] and [12] as an explicitly analytic function of the exterior and interior region of the Schwarzschild black-hole space-time for a finite range of values of the Schwarzschild radial coordinate above and below $r_S = 2M$ ((5) and (8) in the text[1]). Within its range of validity ((6) and (9) in the text) this propagator is a valid approximation to the unknown exact conformal scalar propagator on the Schwarzschild black-hole geometry and has, in fact, been shown in [12] and [13] to be consistent with Hawking radiation. The advantage which this propagator has in its relation to the latter is precisely its explicit reference to the infinity with which the Schwarzschild temporal coordinate labels the event horizon. With $x_1$ labelling a space-time point in the interior region and $x'_1$ signifying the location of the observer in the exterior region, the massless conformal scalar propagator $D(x'_1 - x_1)$ ((12) in the text), obtained in [12] by properly matching the propagators in the interior and the exterior region across the event horizon, explicitly features the periodic Euclidean Schwarzschild temporal coordinate $\tau \to \infty$ on the latter, rendering the effect of infinite time-dilation registered at $x'_1$ manifest. This is precisely the quality which renders $D(x'_1 - x_1)$ semi-classical from the above-stated perspective of the distant observer. It is because, unlike $K(x_1, x'_1)$, the explicit expression $D(x'_1 - x_1)$ centrally involves the event horizon that $D(x'_1 - x_1)$ identically expresses the perspective of the distant static observer in the context of which the emitted particles are in an increasingly semi-classical state when traced back toward the latter. As a direct consequence, he interprets the unique massless conformal scalar propagator $K(x_1, x'_1)$ as the semi-classical propagator $D(x'_1 - x_1)$.

---

[1]Due to their extended nature, the expressions concerning the massless conformal scalar propagator respectively in the exterior and in the interior black-hole geometry are cited in the Introduction by their labels in the text.

For a wavefunction evolving from a surface $C_+$ of constant $r_1$ in the interior region, to a surface of constant $r'_1$ at some $t'_1 = 0$ in the exterior region the implication inherent in $D(x'_1 - x_1)$ is that infinite time-dilation is equivalent to translating the Schwarzschild temporal coordinate $t$ by the amount $-i4\pi M$. Since this is itself equivalent to reflecting the Kruskal coordinates $U$ and $V$ in the origin of the extended Kruskal geometry [2], it follows that the event horizon relates the advance of a mode from $C_+$ to $x'_1 \equiv (0, \vec{r}_1\,')$ to the advance of that mode from white-hole interior-surface $C_-$, which is $C_+$-reflected in the origin of the $U - V$ plane, to $x'_1$. As by dint of time-reversal invariance this relation is precisely that required for thermal emission by the black hole, the novel conclusion is that the effect of infinite time-dilation on the event horizon is the underlying reason for Hawking radiation. The same effect of infinite time-dilation is responsible for the detraction from the thermal spectrum if the condition for a fixed exterior space-time geometry is relaxed. Of the infinitely many detraction values for which infinite time-dilation allows, energy-conservation selects that which corrects the Boltzmann factor $e^{-8\pi ME}$ for a particle of energy $E$ at inverse Hawking temperature $8\pi M$ to $e^{-8\pi E(M-\frac{E}{2})}$. For that matter, the new and central result of this project is that the effect of infinite time-dilation on the event horizon is the underlying reason for Hawking radiation of both, a thermal spectrum and a spectrum which detracts from thermality by the amount which energy-conservation imposes.

A prerequisite for the calculation leading to this conclusion is the determination of the semi-classical propagator's divergence structure. It is in advancing a wavefunction from an interior surface $C_+$ of constant $r_1$ across the event horizon to an exterior surface of constant $r'_1$ at $t'_1 = 0$ that

$$D(x'_1 - x_1) \; ; \; x_1 \equiv (t, r_1, \theta, \phi) \; ; \; x'_1 \equiv (0, \vec{r}_1\,')$$

receives its central physical significance in its relation to Hawking radiation. With these two surfaces respectively considered within the range of this propagator's validity, it will be shown that $D(x'_1 - x_1)$ involves an infinite sequence of null geodesics which spiral an arbitrary number of times near $r = 3M$ and thereby connect the fixed $r_1, \theta_1, \phi_1$-curve in the interior to the fixed space-time point $x'_1$ in the exterior region of the black hole at increasingly large values of $|t - t'_1| = |t|$. This infinite sequence of null geodesics determines the infinite sequence of poles in the general expression $K(x_1, x'_1)$ of the massless conformal scalar propagator in [2]. It will be shown in this respect that although in the present formalism the null geodesics connecting $x_1$ and $x'_1$ do not correspond to poles in $D(x'_1 - x_1)$, the divergence-structure of $D(x'_1 - x_1)$ remains identically the same as that in $K(x_1, x'_1)$ with each such null geodesic generating a corresponding quadratic divergence. This effect is a direct consequence of the specific expression which the unique massless conformal scalar propagator receives in a local frame. The operation of concretizing the general expression $K(x_1, x'_1)$ to the specific expression $D(x'_1 - x_1)$ replaces the poles of the former by the divergent asymptotic behaviour of the latter. As both propagators are identical, their divergence-structure is - albeit different in form - the same.

This project is organized as follows. Section $II$ presents an outline of the massless conformal scalar propagation on the Schwarzschild black-hole space-time established in [11], [12] and [13], contrasts it against $K(x, x')$ in its relation to Hawking radiation in [2] and advances a rigorous derivation of the analyticity properties and behaviour of $D(x'_1 - x_1)$ in transfer-space-related complex plane. Section $III$ identifies the infinite sequence of null geodesics which $D(x'_1 - x_1)$ involves, as well as the infinite sequence of divergences inherent in it and establishes the one-to-one correspondence between them. Section $IV$ advances the rigorous derivation of Hawking radiation and its thermal character as the exclusive consequence of infinite time-dilation on the black hole's event horizon and presents a qualitative argument for the extension of this result to quantum fields of spin greater than zero and to other black-hole geometries. Section $V$ advances an independent derivation of the correction to the thermal spectrum as a consequence of infinite time-dilation on an evolving Schwarzschild event horizon. Finally, Section VI summarises the results and their physical essence.

## II. PROPAGATION ACROSS THE EVENT HORIZON

Before initiating the analysis which will eventuate in the announced result, a brief outline of the massless conformal scalar propagation established in [11], [12] and [13] is in order.

The Schwarzschild metric is:

$$(1) \qquad ds^2 = -(1 - \frac{2M}{r})dt^2 + (1 - \frac{2M}{r})^{-1}dr^2 + r^2(d\theta^2 + sin^2\theta d\phi^2)$$

The analytical extension $\tau = +it$ of the real-time coordinate $t$ to imaginary values results in a Euclidean positive-definite metric for $r > 2M$. The apparent singularity which persists at $r_S = 2M$ can be removed by introducing the new radial coordinate:

$$(2) \qquad \rho = 4M\sqrt{1 - \frac{2M}{r}}$$

Upon replacing:

(3)
$$\beta = 4M$$

the metric in the new coordinates is:

(4)
$$ds^2 = \rho^2(\frac{1}{\beta^2})d\tau^2 + (\frac{4r^2}{\beta^2})^2 d\rho^2 + r^2(d\theta^2 + sin^2\theta d\phi^2)$$

whereupon through the identification of $\frac{\tau}{4M}$ with an angular coordinate of period $2\pi$ the coordinate singularity at $r_S = 2M$ corresponds to the origin $\rho = 0$ of polar coordinates and is thus trivially removed. This procedure results in a complete singularity-free positive-definite Euclidean metric which is periodic in the imaginary-time coordinate $\tau$. The period $8\pi M$ of that coordinate is the underlying cause of the thermal radiation at temperature $T_H = (8\pi M)^{-1}$ emitted by the Schwarzschild black hole.

By imposing the requirement $\rho^2 << \beta^2$ the propagator for a massless conformal scalar field $\Phi$ in the Hartle-Hawking state expressed as an explicitly analytic function of the exterior region of the Schwarzschild black-hole space-time for a certain range of values of the radial variable $r > 2M$ has been established in [11] to be:

$$D(x_2 - x_1) =$$

$$\frac{2}{\beta}\frac{1}{\sqrt{\rho_1\rho_2}}\sum_{l=0}^{\infty}\sum_{m=-l}^{l}Y_{lm}(\theta_2,\phi_2)Y_{lm}^*(\theta_1,\phi_1)\sum_{p=0}^{\infty}e^{i\frac{p}{\beta}(\tau_2-\tau_1)}\int_{u_0[p]}^{\infty}du\frac{cos\left[\frac{\pi}{4\beta}(4u+2p+3)(\rho_2-\rho_1)\right]}{\pi^2 u^2 + 4(l^2+l+1)}$$

$$-\frac{2}{\beta^{\frac{3}{2}}}\frac{1}{\sqrt{\rho_1}}\times$$

$$\sum_{l=0}^{\infty}\sum_{m=-l}^{l}\sum_{p=0}^{\infty}\int_{u_0'[p]}^{\infty}du\frac{cos\left[\frac{\pi}{4\beta}(4u+2p+3)(\beta-\rho_1)\right]}{\pi^2 u^2 + 4(l^2+l+1)}\frac{J_p(\frac{2i}{\beta}\sqrt{l^2+l+1}\rho_2)}{J_p(2i\sqrt{l^2+l+1})}e^{i\frac{p}{\beta}(\tau_2-\tau_1)}Y_{lm}(\theta_2,\phi_2)Y_{lm}^*(\theta_1,\phi_1);$$

(5)
$$u_0 >> p \;;\; u_0' >> p \;;\; \frac{\pi u}{\beta}\rho_{2,1} >> p$$

with a range of validity[2] [11]:

(6)
$$0 \leq \rho_i^2 \leq \frac{\beta^2}{100} \;;\; i = 1,2 \implies 2M \leq r \leq 2.050M$$

and on the understanding that the first expression on the right side of (5) is the singular part $D_{as}(x_2-x_1)$, that part which contains the singularity of $D(x_2-x_1)$ at the coincidence space-time limit $x_2 \to x_1$, whereas the expression following $D_{as}(x_2-x_1)$ is the boundary part $D_b(x_2-x_1)$ which enforces the boundary condition of vanishing propagation:

$$D(x_2 - x_1)_{|\rho=\beta} = 0$$

---

[2]The inadvertent oversight in [11] concerning the upper-bound value $2.0050M$ of that range, was corrected to $2.050M$ in [12].

when $x_2$ or $x_1$ is specified on the boundary hypersurface $\rho = \beta$.

In the interior region $r < 2M$ of (1) the analytical extension to imaginary time also involves the analytical extension $\theta \to i\tilde{\theta}$ as a result of which (1) reduces to the negative-definite [14]:

$$ds^2 = -[[\frac{2M}{r} - 1]d\tau^2 + \frac{1}{\frac{2M}{r} - 1}dr^2 + r^2(d\tilde{\theta}^2 + sinh^2\tilde{\theta}d\phi^2)] \tag{7}$$

The massless conformal scalar propagator $D^{(int)}(x_2 - x_1)$ has been established in the interior region of the Schwarzschild black hole for a certain range of values of $r < r_S$ through the analytical extension of (5) [12]. With $\tau = \pm i\xi$ and $\rho = \pm i\zeta$ the singular part of that propagator is:

$$D^{(int)}_{as}(x_2 - x_1) = \frac{i}{\beta} \frac{e^{-\frac{3\pi}{4\beta}(\zeta_2 - \zeta_1)}}{\sqrt{\zeta_1\zeta_2}} \int_{\epsilon \to 0}^{\infty} d|\tilde{p}| \frac{1}{1 - e^{-2\pi|\tilde{p}|}} e^{i\frac{|\tilde{p}|}{\beta}(\xi_2 - \xi_1)} e^{i\frac{\pi}{\beta}\left[\frac{|\tilde{p}|}{2} + u_0[|\tilde{p}|]\right](\zeta_2 - \zeta_1)} \times$$

$$\sum_{l=0}^{\infty} \sum_{m=-l}^{l} Y_{lm}(i\tilde{\theta}_2, \phi_2) Y^*_{lm}(i\tilde{\theta}_1, \phi_1) \int_0^{\infty} dw \frac{e^{-i\frac{\pi}{\beta}w(\zeta_2 - \zeta_1)}}{\pi^2(u_0[|\tilde{p}|] - w)^2 - 4(l^2 + l + 1)}$$

$$+\frac{i}{\beta} \frac{e^{\frac{3\pi}{4\beta}(\zeta_2 - \zeta_1)}}{\sqrt{\zeta_1\zeta_2}} \int_{\epsilon \to 0}^{\infty} d|\tilde{p}| \frac{1}{1 - e^{-2\pi|\tilde{p}|}} e^{i\frac{|\tilde{p}|}{\beta}(\xi_2 - \xi_1)} e^{i\frac{\pi}{\beta}\left[\frac{|\tilde{p}|}{2} + u_0[|\tilde{p}|]\right](\zeta_2 - \zeta_1)} \times$$

$$\sum_{l=0}^{\infty} \sum_{m=-l}^{l} Y_{lm}(i\tilde{\theta}_2, \phi_2) Y^*_{lm}(i\tilde{\theta}_1, \phi_1) \int_0^{\infty} dw \frac{e^{i\frac{\pi}{\beta}w(\zeta_2 - \zeta_1)}}{\pi^2(u_0[|\tilde{p}|] + w)^2 - 4(l^2 + l + 1)}$$

$$+\frac{i}{\beta}\frac{1}{2\pi} \frac{e^{i\frac{\pi}{\beta}u_0[0](\zeta_2 - \zeta_1)}}{\sqrt{\zeta_1\zeta_2}} \int_{\frac{\pi}{2}}^{\frac{3\pi}{2}} d\theta e^{i\theta} e^{i\frac{3\pi}{4\beta}e^{-i\theta}(\zeta_2 - \zeta_1)} \times$$

$$\sum_{l=0}^{\infty} \sum_{m=-l}^{l} Y_{lm}(i\tilde{\theta}_2, \phi_2) Y^*_{lm}(i\tilde{\theta}_1, \phi_1) \int_0^{\infty} dw \frac{e^{-\frac{\pi}{\beta}we^{-i\theta}(\zeta_2 - \zeta_1)}}{\pi^2(u_0[0]e^{i\theta} + iw)^2 + 4(l^2 + l + 1)}$$

$$-\frac{i}{\beta} \frac{e^{\frac{3\pi}{4\beta}(\zeta_2 - \zeta_1)}}{\sqrt{\zeta_1\zeta_2}} \int_{\epsilon \to 0}^{\infty} d|\tilde{p}| \frac{1}{1 - e^{-2\pi|\tilde{p}|}} e^{i\frac{|\tilde{p}|}{\beta}(\xi_2 - \xi_1)} e^{-i\frac{\pi}{\beta}\left[\frac{|\tilde{p}|}{2} + u_0[|\tilde{p}|]\right](\zeta_2 - \zeta_1)} \times$$

$$\sum_{l=0}^{\infty} \sum_{m=-l}^{l} Y_{lm}(i\tilde{\theta}_2, \phi_2) Y^*_{lm}(i\tilde{\theta}_1, \phi_1) \int_0^{\infty} dw \frac{e^{-i\frac{\pi}{\beta}w(\zeta_2 - \zeta_1)}}{\pi^2(u_0[|\tilde{p}|] + w)^2 - 4(l^2 + l + 1)}$$

$$-\frac{i}{\beta} \frac{e^{-\frac{3\pi}{4\beta}(\zeta_2 - \zeta_1)}}{\sqrt{\zeta_1\zeta_2}} \int_{\epsilon \to 0}^{\infty} d|\tilde{p}| \frac{1}{1 - e^{-2\pi|\tilde{p}|}} e^{i\frac{|\tilde{p}|}{\beta}(\xi_2 - \xi_1)} e^{-i\frac{\pi}{\beta}\left[\frac{|\tilde{p}|}{2} + u_0[|\tilde{p}|]\right](\zeta_2 - \zeta_1)} \times$$

$$\sum_{l=0}^{\infty} \sum_{m=-l}^{l} Y_{lm}(i\tilde{\theta}_2, \phi_2) Y^*_{lm}(i\tilde{\theta}_1, \phi_1) \int_0^{\infty} dw \frac{e^{i\frac{\pi}{\beta}w(\zeta_2 - \zeta_1)}}{\pi^2(u_0[|\tilde{p}|] - w)^2 - 4(l^2 + l + 1)}$$

$$-\frac{i}{\beta}\frac{1}{2\pi} \frac{e^{-i\frac{\pi}{\beta}u_0[0](\zeta_2 - \zeta_1)}}{\sqrt{\zeta_1\zeta_2}} \int_{\frac{\pi}{2}}^{\frac{3\pi}{2}} d\theta e^{i\theta} e^{-i\frac{3\pi}{4\beta}e^{-i\theta}(\zeta_2 - \zeta_1)} \times$$

$$
(8) \quad \sum_{l=0}^{\infty} \sum_{m=-l}^{l} Y_{lm}(i\tilde{\theta}_2, \phi_2) Y_{lm}^*(i\tilde{\theta}_1, \phi_1) \int_0^{\infty} dw \frac{e^{-\frac{\pi}{\beta} w e^{-i\theta}(\zeta_2 - \zeta_1)}}{\pi^2 (u_0[0] e^{i\theta} - iw)^2 + 4(l^2 + l + 1)}
$$

with a range of validity for $D^{(int)}(x_2 - x_1)$ [12]:

$$
(9) \quad 0 \leq |\rho_{int}|^2 \leq \frac{\beta^2}{100} \implies 1.980 M \leq r \leq 2M
$$

In passing, attention is invited to the fact that the notation $i\tilde{\theta}$ in the angular sector of $D_{as}^{(int)}(x_2 - x_1)$ stresses the stated analytical extension $\theta \to i\tilde{\theta}$, yet signifies the real positive-definite magnitude $\tilde{\theta}$. This can be directly elicited from the interior metric in (7). For that matter, $Y_{lm}(i\tilde{\theta}, \phi)$ really signifies $Y_{lm}(\tilde{\theta}, \phi)$.

As stated in some detail in the Introduction, each particle emitted by the black hole is in an increasingly semi-classical state of very high energy when traced back toward the event horizon. In turn, at the semi-clasical limit $\hbar \to 0$ the dominant contribution to the propagator $D(x'_1 - x_1)$ which describes the amplitude that a particle transit from a point $x_1$ in interior region $II$ to a point $x'_1$ in exterior region $I$ where it is registered in a local frame (Fig.1) stems from eigenvalues and eigenstates [11]:

$$
-\Box \phi_m = \lambda_m \phi_m
$$

of very high order. The implication of semi-classicality is that the transition amplitude reduces in the stated frame to [13]:

$$
(10) \quad D(x'_1 - x_1) = e^{-S_0[\Phi_0, \nabla_\alpha \Phi_0] + O(\hbar)}
$$

with $S_0[\Phi_0, \nabla_\alpha \Phi_0]$ being the classical action and with $O(\hbar)$ describing higher-loop corrections to it, whereas the implication of very high energy is that the entire transition amplitude reduces to merely the asymptotic part of the propagator [13]:

$$
(11) \quad D(x'_1 - x_1) = D_{as}(x'_1 - x_1)
$$

This latter aspect of quantum propagation is the reason the boundary part of $D^{(int)}(x_2 - x_1)$ which has been derived in [12] is not relevant to the present analysis.

The amplitude that an excitation of a conformal scalar field transit from a space-time point $x_1 \equiv (t_1, r_1, \tilde{\theta}_1, \phi_1)$ in the interior region to a space-time point $x'_1 \equiv (t'_1, r'_1, \theta'_1, \phi'_1)$ in the exterior region of the Schwarzschild black-hole geometry is expressed by the propagator [12], [13]:

$$
D(x'_1 - x_1) = D_{as}(x'_1 - x_1) =
$$

$$
(2M)^2 \int_{-1}^{1} d\cos\theta_2 \int_0^{2\pi} d\phi_2 \Big[ \frac{i}{8\pi M} \frac{1}{\sqrt{\zeta_1}} \sum_{l=0}^{\infty} \sum_{m=-l}^{l} Y_{lm}(\theta_2, \phi_2) Y_{lm}^*(\tilde{\theta}_1, \phi_1) \times
$$

$$[e^{-i\frac{\pi}{4M}u_0[0]\zeta_1}\int_{\frac{\pi}{2}}^{\frac{3\pi}{2}}d\theta e^{i\theta}e^{-i\frac{3\pi}{16M}e^{-i\theta}\zeta_1}\int_0^\infty d\omega \frac{e^{\frac{\pi}{4M}we^{-i\theta}\zeta_1}}{\pi^2(u_0[0]e^{i\theta}+i\omega)^2+4(l^2+l+1)}-$$

$$e^{i\frac{\pi}{4M}u_0[0]\zeta_1}\int_{\frac{\pi}{2}}^{\frac{3\pi}{2}}d\theta e^{i\theta}e^{i\frac{3\pi}{16M}e^{-i\theta}\zeta_1}\int_0^\infty d\omega \frac{e^{\frac{\pi}{4M}we^{-i\theta}\zeta_1}}{\pi^2(u_0[0]e^{i\theta}-i\omega)^2+4(l^2+l+1)}]\times$$

(12) $$\frac{1}{M^2}\frac{1}{32\pi^2}\frac{1}{\sqrt{\rho'_1}}\sum_{k=0}^\infty\sum_{p>p_0>0}^\infty e^{i\frac{p}{4M}n(8\pi M)}e^{-i\frac{p}{4M}\tau'_1}\frac{2k+1}{p}P_k(\cos\gamma)]$$

with:

(i) $\qquad \cos\gamma = \cos\theta_2\cos\theta'_1 + \sin\theta_2\sin\theta'_1\cos(\phi_2-\phi'_1)$ ,

(ii) the integration over $\cos\theta_2$ and $\phi_2$ signifying a superposition over a complete set of states defined on the future event horizon of the extended Kruskal geometry (Fig.1) and

(iii) $n$ being the number of periods $8\pi M$ of the Euclidean Schwarzschild temporal coordinate $\tau_2$ in view of the fact that in the Euclidean sector of the Schwarzschild metric, respectively expressed by (4) and (7), time is periodic with period $8\pi M$. Crucial to the ensuing analysis in this respect is the fact that at the value $r_S = 2M$ of the Schwarzschild radial coordinate, the infinite time-dilation on the future event horizon implies that:

(13) $$\tau_2 \to \infty \Longrightarrow n \to \infty$$

This concludes the outline of the massless conformal scalar propagation established in [11], [12] and [13].

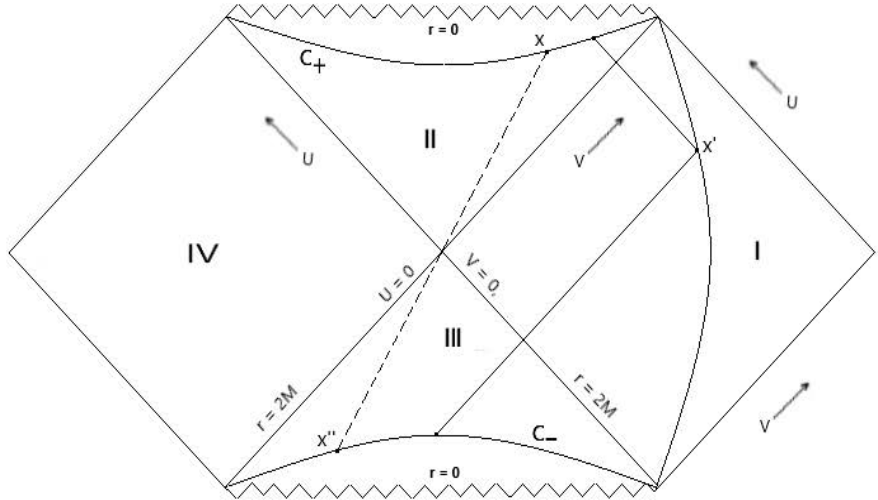

Figure 1. A reproduction of the *Penrose diagram for the Schwarzschild geometry* in [2].

The amplitude that the Schwarzschild black hole emit particles which are detected in a mode of positive energy $E$ by an observer in region $I$ is expressed by the evolution of the wavefunction from a space-like surface $C_+$ of constant $r = r_1$ in region $II$ of the extended Kruskal geometry to the world line of the observer (Fig.1). Central to that evolution is the propagator [2]:

$$K(x, x') = -\frac{i}{4\pi^2} \frac{1}{s(x, x') + i\epsilon} \tag{14}$$

with $s(x, x')$ being the square of the interval between points $x \equiv (t, r_1, \tilde{\theta}, \phi)$ on $C_+$ and $x' \equiv (t', r', \theta', \phi')$ in region $I$ of the Schwarzschild black-hole space-time. This propagator involves an infinite sequence of poles. If the arbitrary point $x'$ is concretized to the fixed point $x'_1 \equiv (t'_1, r'_1, \theta'_1, \phi'_1)$ on the world line of the observer, these poles correspond to the null geodesics which spiral an arbitrary number of times near $r = 3M$ and thus connect the fixed $r_1$-surface $C_+$ in region $II$ (represented in the Penrose diagram by the fixed $r_1, \tilde{\theta}_1, \phi_1$-curve $C_+$) to the fixed $x'_1$ at increasingly large values of $|t - t'_1|$ (Fig.2).

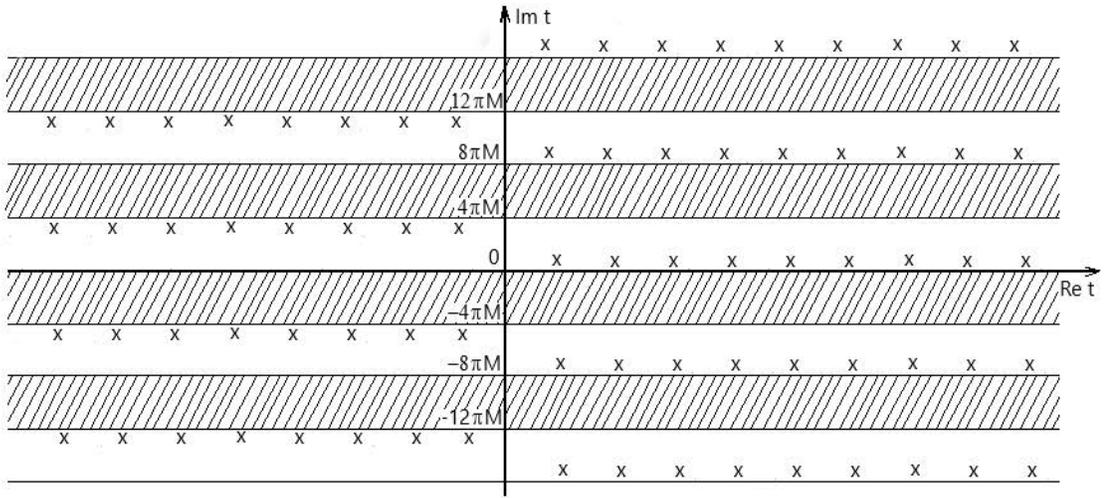

Figure 2. A reproduction of *the analytic structure of the propagator in the complex t-plane* in [2]. The shaded regions are the regions of analyticity of $K(x, x')$ in $t$. The crosses immediately above the real $t$-axis locate the singularities which correspond to the real null geodesics connecting the fixed point $x'_1$ in region $I$ to the fixed $r_1, \tilde{\theta}_1, \phi_1$-curve $C_+$ in region $II$. The crosses immediately below the $Imt = -4\pi M$-axis locate the singularities which correspond to the real null geodesics connecting $x'_1$ to the fixed $r_1, \tilde{\theta}_1, \phi_1$-curve $C_-$ in region $III$. The singularities at other values of $Imt$ are duplicates of these as a consequence of the periodicity of the propagator in $Imt$ with period $8\pi M$.

The technique, introduced by Hartle and Hawking in [2], which derives the black-hole radiation from the quantum-mechanical perspective of particle-propagation is predicated on the premise that the amplitude a black hole emit a particle which is detected in a mode of positive energy $E$ at fixed space-time point $x'_1 \equiv (t'_1 = 0, r'_1, \theta'_1, \phi'_1)$ in region $I$ - with $t'_1 = 0$ being a convenient choice - is:

$$\mathcal{E}_E(\vec{r}_1', \vec{r}_1) = \int_{-\infty}^{+\infty} dt\, e^{-iEt} K(0, \vec{r}_1'; t, \vec{r}_1) = \int_{-\infty}^{+\infty} dt\, e^{-iEt} K(t, \vec{r}_1; 0, \vec{r}_1') \tag{15}$$

where the identification $x_1 \equiv x \equiv (t, r_1, \tilde{\theta}, \phi) \equiv (t, \vec{r}_1)$ for an arbitrary space-time point on surface $C_+$ (fixed $r_1, \tilde{\theta}_1, \phi_1$-curve in region $II$ (Fig.1)) will henceforth be used in order to bring the notation in (15) in line with that in (12). The premise expressed in (15) is essentially an extension of the quantum-mechanical statement:

$$\langle \vec{x}\,', t' | \psi \rangle = \int d^3x \langle \vec{x}\,', t' | \vec{x}, t \rangle \langle \vec{x}, t | \psi \rangle \tag{16}$$

to the Schwarzschild black-hole space-time on the understanding that:

(i) the integration in (15) extends over the space-like coordinate $t$ on the fixed surface $C_+$ in region $II$.

(ii) the transition amplitude $\langle \vec{r}_1', 0 | \vec{r}_1, t \rangle$ is identified as the propagator $K(0, \vec{r}_1'; t, \vec{r}_1)$.

(iii) it is because $t$ is space-like that $E$ must not in region $II$ be interpreted as the energy.

Central to the derivation of Hawking radiation and its thermal spectrum in [2] is the equality between the amplitude in (15) and the amplitude obtained through the distortion of the contour of $t$-integration by an amount $-i4\pi M$. That equality emerges as a consequence of the analyticity properties of the propagator and its behavior in the $t$-complex plane demonstrated in Fig.2. Such operation as the stated distortion signifies is, in turn, equivalent to reflecting the Kruskal coordinates $U$ and $V$ in the origin so that point $x$ on surface $C_+$ in region $II$ is reflected to point $x''$ on surface $C_-$ in region $III$. For that matter, the amplitude in (15) receives the equivalent interpretation of $e^{-E4\pi M}$ times the amplitude a particle propagate from surface $C_-$, the $C_+$-reflection in the origin of the $U-V$ plane, to point $x'$ on the observer's world line in region $I$ to be registered in a mode of positive energy $E$ (Fig.1):

$$\mathcal{E}_E(\vec{r}_1', \vec{r}_1) = e^{-E4\pi M} \int_{-\infty}^{+\infty} dt e^{-iEt} K(t - i4\pi M, \vec{r}_1\ ;\ 0, \vec{r}_1') =$$

$$e^{-(8\pi M)\frac{E}{2}} \int_{-\infty}^{+\infty} dt e^{-iEt} K(t - i4\pi M, \vec{r}_1\ ;\ 0, \vec{r}_1') \tag{17}$$

By time-reversal invariance, the modulus squared of this latter amplitude is exactly equal to the modulus squared of the amplitude the black hole absorb a particle which emanates from $x_1' \equiv (0, \vec{r}_1')$ in a mode of positive-definite energy $E$ and eventuates at $C_+$ in a mode which corresponds to the same energy. This connection between emission and absorption reveals that a black hole emits particles with a thermal spectrum at temperture $T_H = \frac{1}{8\pi M}$.

In order to examine the implications of (15) to scalar propagation from the perspective of the observer in region $I$, it is necessary to assess the consequences which the infinite time dilation on the event horizon has on that propagation. To that effect, replacing the general expression $K(x_1, x_1')$ of the propagator in (15) by the explicit function of space-time $D(x_1' - x_1)$ in (12) yields:

$$\mathcal{E}_E(\vec{r}_1', \vec{r}_1) = \int_{-\infty}^{+\infty} dt e^{-iEt} D(x_1' - x_1) \tag{18}$$

with $x'_1 \equiv (0, \vec{r}'_1)$ in region $I$ and $x_1 \equiv (t, \vec{r}_1)$ on $C_+$ in region $II$ respectively considered within the stated range of validity in (6) and (9). The evaluation of the amplitude for particle emission necessitates the observation that, as a consequence of (13), $D(x'_1 - x_1)$ involves an infinite sequence of null geodesics labeled by $n, n+1, n+2, ....$ They are the stated null geodesics which spiral an arbitrary number of times near $r = 3M$ and thus connect the fixed $r_1, \tilde{\theta}_1, \phi_1$-curve $C_+$ in region $II$ to the fixed space-time point $x'_1$ in region $I$ at increasingly large values of $|t|$. They correspond, for that matter, to the infinite sequence of poles which the propagator $K(x_1, x'_1)$ has in the complex $t$-plane. Careful examination of the analytical behavior of $D(x' - x)$ will reveal this infinite sequence.

The propagator as an explicit function of space-time in (12) in the context of (6) and (9) also involves infinitely many poles, now in the $p$-complex plane as is evident in the infinite series over $p$. The representation of this infinite series as a contour integral which explicitly features that set of infinitely many poles is necessary for the exploration of scalar propagation [13]. For that matter, upon analytically extending $p \in R$ to $\tilde{p} \in C$ the residue theorem recasts the infinite series over $p$ in (12) as:

$$(19) \quad \sum_{p>p_0>0}^{\infty} \frac{1}{p} e^{i\frac{p}{4M}n(8\pi M)} e^{-i\frac{p}{4M}\tau'_1} = \frac{1}{2i} e^{i\frac{p_0+1}{4M}n(8\pi M)} \oint_{c_{\tilde{p}}} d\tilde{p} \frac{(-1)^{\tilde{p}}}{sin(\pi\tilde{p})} \frac{1}{\tilde{p}} e^{i\frac{\tilde{p}-(p_0+1)}{4M}n(8\pi M)} e^{-i\frac{\tilde{p}}{4M}\tau'_1}$$

with the contour integral considered over the infinite contour $c_{\tilde{p}}$ in Fig.3 which encompasses all integers bigger than some $p_0 \geq 1$.

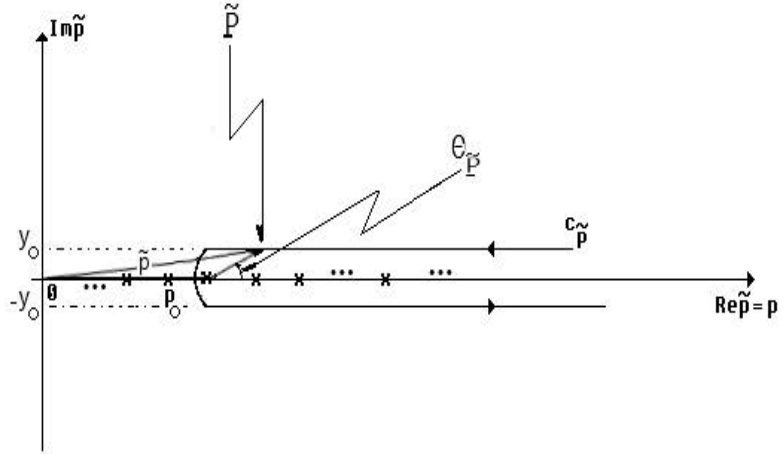

Figure 3. The infinite contour $c_{\tilde{p}}$ with $\tilde{P} = \tilde{p} - p_0$.

Setting for convenience $\tilde{P} = \tilde{p} - (p_0+1)$, or simply (since $p_0 > 0$ is arbitrary) $\tilde{P} = \tilde{p} - p_0$, (19) becomes:

$$\sum_{p>p_0>0}^{\infty} \frac{1}{p} e^{i\frac{p}{4M}n(8\pi M)} e^{-i\frac{p}{4M}\tau_1'} =$$

(20) $\quad \frac{1}{2i} e^{i\frac{p_0}{4M}n(8\pi M)} e^{-i\frac{p_0}{4M}\tau_1'} e^{i\pi p_0} \oint_{c_{\tilde{p}}} d\tilde{P} \frac{(-1)^{\tilde{P}}}{sin(\pi\tilde{P}+\pi p_0)} \frac{1}{\tilde{P}+p_0} e^{i\frac{\tilde{P}}{4M}n(8\pi M)} e^{-i\frac{\tilde{P}}{4M}\tau_1'}$

With $\tilde{P} = |\tilde{P}|e^{i\theta_{\tilde{P}}}$, $x = Re\tilde{P} + p_0$ and $y = Im\tilde{P}$, the contribution to the contour integral along the segment $x \in (\infty, p_0] \implies \theta_{\tilde{P}} \in (0, \frac{\pi}{2}]$ ; $y = y_0$ of $c_{\tilde{p}}$ is:

$$e^{-i\pi p_0} e^{-\pi y_0} e^{-i\frac{p_0}{4M}n(8\pi M)} e^{-\frac{y_0}{4M}n(8\pi M)} \times$$

(21) $\quad \int_{\infty+iy_0}^{p_0+iy_0} d(x+iy_0) \frac{e^{i\pi x}}{sin[\pi(x+iy_0)]} \frac{1}{x+iy_0} e^{i\frac{x}{4M}n(8\pi M)} e^{-i\frac{|\tilde{P}|}{4M}t_1'}$

on the understanding that:

(i) $\tilde{P} = |\tilde{P}|e^{i\theta_{\tilde{P}}} \implies \tau_1' = t_1' e^{-i\theta_{\tilde{P}}}$, which follows from the fact that an analytical extension in transfer space is offset in the propagator by a corresponding analytical extension of the associated coordinate variable [12],

(ii) $\tau_2 = n(8\pi M)$ cannot be analytically continued to the real $t$-axis on account of (13).

For that matter, as a consequence of $n \to \infty$ on the event horizon, the integrand oscillates with infinite frequency, causing the contribution in (21) to vanish (all the more so on account of $e^{-\frac{y_0}{4M}n(8\pi M)}$).

Accordingly, the contribution to the contour integral along the segment $x \in [p_0, \infty) \implies \theta_{\tilde{P}} \in [\frac{3\pi}{2}, 2\pi)$ ; $y = -y_0$ of $c_{\tilde{p}}$ is:

$$e^{-i\pi p_0} e^{\pi y_0} e^{-i\frac{p_0}{4M}n(8\pi M)} e^{\frac{y_0}{4M}n(8\pi M)} \times$$

(22) $\quad \int_{p_0-iy_0}^{\infty-iy_0} d(x-iy_0) \frac{e^{i\pi x}}{sin[\pi(x-iy_0)]} \frac{1}{x-iy_0} e^{i\frac{x}{4M}n(8\pi M)} e^{-i\frac{|\tilde{P}|}{4M}t_1'}$

This integral also vanishes on account of $n \to \infty$.

The contribution to the contour integral along the remaining segment of $c_{\tilde{p}}$ specified by $|\tilde{P}| = y_0$ and $\theta_{\tilde{P}} \in (\frac{\pi}{2}, \frac{3\pi}{2})$ is:

(23) $\quad i|\tilde{P}| \int_{\frac{\pi}{2}}^{\frac{3\pi}{2}} d\theta_{\tilde{P}} e^{i\theta_{\tilde{P}}} \frac{e^{i\pi|\tilde{P}|e^{i\theta_{\tilde{P}}}}}{sin[\pi(|\tilde{P}|e^{i\theta_{\tilde{P}}}+p_0)]} \frac{e^{i\frac{|\tilde{P}|}{4M}ncos\theta_{\tilde{P}}(8\pi M)}}{|\tilde{P}|e^{i\theta_{\tilde{P}}}+p_0} e^{-\frac{|\tilde{P}|}{4M}nsin\theta_{\tilde{P}}(8\pi M)} e^{-i\frac{|\tilde{P}|}{4M}t_1'}$

Again, on the event horizon ($n \to \infty$) the factor $e^{i\frac{|\tilde{P}|}{4M}ncos\theta_{\tilde{P}}(8\pi M)}$ causes the integral to vanish unless $e^{i\theta_{\tilde{P}}}$ is, for all intents and purposes, a constant. In turn, the only contribution to (20) stems from that segment of $c_{\tilde{p}}$'s semi-circle along which $cos\theta_{\tilde{P}} \sim -1$:

$$\sum_{p>p_0>0} \frac{1}{p} e^{i\frac{p}{4M}n(8\pi M)} e^{-i\frac{p}{4M}\tau'_1} =$$

$$(24) \quad e^{-8\pi M w'}\left[Ce^{i\frac{p_0}{4M}n(8\pi M)}e^{\frac{p_0}{4M}t'_1}e^{i\pi p_0}|\tilde{P}|\frac{e^{-i\pi|\tilde{P}|}}{sin(-\pi|\tilde{P}|+\pi p_0)}\frac{e^{-i\frac{|\tilde{P}|}{4M}n(8\pi M)}}{-|\tilde{P}|+p_0}e^{-i\frac{|\tilde{P}|}{4M}t'_1}\right]$$

where $e^{-i\frac{p_0}{4M}\tau'_1} = e^{-i\frac{p_0}{4M}(it'_1)} = e^{\frac{p_0}{4M}t'_1}$, $C$ is a real constant and

$$(25) \quad w' = \frac{|\tilde{P}|}{4M}(nsin\theta_{\tilde{P}})$$

Replacing (24) in (12) yields[3]

$$D(x'_1 - x_1) = e^{-8\pi M w'}\left[Ce^{i\frac{p_0}{4M}n(8\pi M)}e^{\frac{p_0}{4M}t'_1}e^{i\pi p_0}|\tilde{P}|\frac{e^{-i\pi|\tilde{P}|}}{sin(-\pi|\tilde{P}|+\pi p_0)}\frac{e^{-i\frac{|\tilde{P}|}{4M}n(8\pi M)}}{-|\tilde{P}|+p_0}e^{-i\frac{|\tilde{P}|}{4M}t'_1}\right.\times$$

$$\frac{i}{64\pi^3 M}\frac{1}{\sqrt{\zeta_1}}\frac{1}{\sqrt{\rho'_1}}\int_{-1}^{1}dcos\theta_2\int_{0}^{2\pi}d\phi_2\sum_{l=0}^{\infty}\sum_{m=-l}^{l}Y_{lm}(\theta_2,\phi_2)Y^*_{lm}(\tilde{\theta}_1,\phi_1)\times$$

$$[e^{-i\frac{\pi}{4M}u_0[0]\zeta_1}\int_{\frac{\pi}{2}}^{\frac{3\pi}{2}}d\theta e^{i\theta}e^{-i\frac{3\pi}{16M}e^{-i\theta}\zeta_1}\int_{0}^{\infty}d\omega\frac{e^{\frac{\pi}{4M}we^{-i\theta}\zeta_1}}{\pi^2(u_0[0]e^{i\theta}+i\omega)^2+4(l^2+l+1)} -$$

$$e^{i\frac{\pi}{4M}u_0[0]\zeta_1}\int_{\frac{\pi}{2}}^{\frac{3\pi}{2}}d\theta e^{i\theta}e^{i\frac{3\pi}{16M}e^{-i\theta}\zeta_1}\int_{0}^{\infty}d\omega\frac{e^{\frac{\pi}{4M}we^{-i\theta}\zeta_1}}{\pi^2(u_0[0]e^{i\theta}-i\omega)^2+4(l^2+l+1)}]\times$$

$$(26) \quad \sum_{k=0}^{\infty}(2k+1)P_k(cos\gamma)]$$

with $\rho'_1$ given by (2) and $\zeta_1 = e^{i\theta_{\tilde{P}}}\rho_1$ [12]. The procedure which eventuated in (26) is a comprehensive rigorous derivation of the corresponding result in [13].

In respect to (25), it must be stressed that the positive orientation of the contour $c_{\tilde{P}}$ constitutes the boundary condition which ensures that the emitted particles are positive-energy states which propagate forward in time [13]. In view of the importance which this aspect of the theory has, a reproduction of the relevant derivation briefly follows.

As a consequence of $n \to \infty$, the limit $lim_{\theta_{\tilde{P}} \to \pi}nsin\theta_{\tilde{P}}$ is a real constant. Whether this constant is positive or negative is detemined by the fact that for

$$\pi - \epsilon < \theta_{\tilde{P}} < \pi + \epsilon \Longrightarrow sin\theta_{\tilde{P}} \sim \pi - \theta_{\tilde{P}} \equiv \chi \in [-\epsilon, \epsilon]$$

the limit $n \to \infty$ coincides with $lim_{\chi \to 0^+}\frac{c}{\chi}$ ; $c > 0$. In turn:

---

[3]Some caution is required at $\theta \in (\frac{3\pi}{2} - \epsilon, \frac{3\pi}{2})$ on account of an apparent fixed-frequency - infinite-amplitude oscillation in that domain. The replacement of $cos\theta$ and $sin\theta$ respectively by $cos\theta + iv$ and $sin\theta + iv$, $0 < v << 1$ confirms the qualitative expectation of a vanishing integral.

$$lim_{\theta_{\tilde{P}} \to \pi^-} n sin\theta_{\tilde{P}} = lim_{\chi \to 0^+} \frac{c}{\chi}\chi = c > 0$$

with the direct implication that:

$$lim_{\theta_{\tilde{P}} \to \pi} n sin\theta_{\tilde{P}} = c > 0$$

since this real positive constant has already been attained as integration over $\theta_{\tilde{P}}$ in (23) advances past $\pi$.

Clearly, a negative (clockwise) orientation of the contour $c_{\tilde{P}}$ would accordingly result in a negative real constant at $\theta_{\tilde{P}} \to -\pi$. It follows, for that matter, that the demand for a positive-definite $w'$ in (25) imposes the stated boundary condition to the effect that integration in (19) advance over the positively-oriented contour $c_{\tilde{P}}$ in Fig.3. Since, as qualitatively stated in [13] and will be rigorously derived in the fourth section, $w' = \frac{E}{2}$ with $E > 0$ being the energy of the emitted particle, the physical demand that the emitted particles be states of positive-definite energy which propagate forward in time imposes the same boundary condition.

### III. THE DIVERGENCE STRUCTURE

As, pursuant to (10), the propagator in (26) has a manifestly semi-classical character, it is expected to involve the afore-stated infinite sequence of null geodesics in the dominant $e^{-(8\pi M)w'}$-factor. That this is indeed the case follows as a direct consequence of the fact that $w'$ in (25) is, at the limit $n \to \infty$, a finite unspecified constant. The sequence of infinitely many divergences in one-to-one correspondence with the sequence of infinitely many null geodesics is now inherent in (26). The objective in this section is to identify both sequences and establish the one-to-one correspondence between them.

Since the infinite series in (26) has been obtained in [13] through use of the addition theorem for spherical harmonics:

(27) $$\sum_{l=0}^{\infty} \sum_{m=-l}^{l} Y_{lm}(\theta_2, \phi_2) Y_{lm}^*(\theta_1', \phi_1') = \delta(cos\theta_2 - cos\theta_1')\delta(\phi_2 - \phi_1')$$

and subsequent application of the identity:

(28) $$\delta(cos\theta_2 - cos\theta_1')\delta(\phi_2 - \phi_1') = \frac{1}{4\pi} \sum_{k=0}^{\infty} (2k+1) P_k(cos\gamma)$$

with:

(29) $$cos\gamma = cos\theta_1' cos\theta_2 + sin\theta_1' sin\theta_2 cos(\phi_1' - \phi_2)$$

it follows that:

$$\int_{-1}^{1} d\cos\theta_2 \int_{0}^{2\pi} d\phi_2 \sum_{l=0}^{\infty} \sum_{m=-l}^{l} Y_{lm}(\theta_2, \phi_2) Y_{lm}^*(\tilde{\theta}_1, \phi_1) [f_1(l) - f_2(l)] \sum_{k=0}^{\infty} (2k+1) P_k(\cos\gamma) =$$

$$4\pi \int_{-1}^{1} d\cos\theta_2 \int_{0}^{2\pi} d\phi_2 \delta(\cos\theta_2 - \cos\theta_1') \delta(\phi_2 - \phi_1') \times$$

(30)
$$\sum_{l=0}^{\infty} \sum_{m=-l}^{l} Y_{lm}(\theta_2, \phi_2) Y_{lm}^*(\tilde{\theta}_1, \phi_1) [f_1(l) - f_2(l)]$$

with $[f_1(l) - f_2(l)]$ being the extended $l$-dependent mathematical expression multiplying $\sum_{l=0}^{\infty} \sum_{m=-l}^{l} Y_{lm}(\theta_2, \phi_2) Y_{lm}^*(\tilde{\theta}_1, \phi_1)$ in (26).

In line with the notation used in [2], the real positive-definite coordinate angle $\tilde{\theta}_1$ labelling space-time points on $C_+$ will, henceforth, be denoted by $\theta_1$. For that matter, (30) yields:

$$\int_{-1}^{1} d\cos\theta_2 \int_{0}^{2\pi} d\phi_2 \sum_{l=0}^{\infty} \sum_{m=-l}^{l} Y_{lm}(\theta_2, \phi_2) Y_{lm}^*(\theta_1, \phi_1) [f_1(l) - f_2(l)] \sum_{k=0}^{\infty} (2k+1) P_k(\cos\gamma) =$$

$$4\pi \sum_{l=0}^{\infty} \sum_{m=-l}^{l} Y_{lm}(\theta_1' + 2\pi n', \phi_1') Y_{lm}^*(\theta_1, \phi_1) [f_1(l) - f_2(l)] \; ; \; n' = 0, 1, 2, ...$$

in view of which (26) implies:

$$D(x_1' - x_1) = e^{-8\pi M w'} \Big[ C e^{i\frac{p_0}{4M} n(8\pi M)} e^{\frac{p_0}{4M} t_1'} e^{i\pi p_0} |\tilde{P}| \frac{e^{-i\pi|\tilde{P}|}}{\sin(-\pi|\tilde{P}| + \pi p_0)} \frac{e^{-i\frac{|\tilde{P}|}{4M} n(8\pi M)}}{-|\tilde{P}| + p_0} e^{-i\frac{|\tilde{P}|}{4M} t_1'} \times$$

$$\frac{i}{16\pi^2 M} \frac{1}{\sqrt{\zeta_1}} \frac{1}{\sqrt{\rho_1'}} \sum_{l=0}^{\infty} \sum_{m=-l}^{l} Y_{lm}(\theta_1' + 2\pi n', \phi_1') Y_{lm}^*(\theta_1, \phi_1) \times$$

$$[e^{-i\frac{\pi}{4M} u_0[0]\zeta_1} \int_{\frac{\pi}{2}}^{\frac{3\pi}{2}} d\theta e^{i\theta} e^{-i\frac{3\pi}{16M} e^{-i\theta} \zeta_1} \int_{0}^{\infty} d\omega \frac{e^{\frac{\pi}{4M} w e^{-i\theta} \zeta_1}}{\pi^2 (u_0[0] e^{i\theta} + i\omega)^2 + 4(l^2 + l + 1)} -$$

(31)
$$e^{i\frac{\pi}{4M} u_0[0]\zeta_1} \int_{\frac{\pi}{2}}^{\frac{3\pi}{2}} d\theta e^{i\theta} e^{i\frac{3\pi}{16M} e^{-i\theta} \zeta_1} \int_{0}^{\infty} d\omega \frac{e^{\frac{\pi}{4M} w e^{-i\theta} \zeta_1}}{\pi^2 (u_0[0] e^{i\theta} - i\omega)^2 + 4(l^2 + l + 1)} \Big]\Big]$$

It becomes obvious now that in view of:

$$\sum_{l=0}^{\infty} \sum_{m=-l}^{l} Y_{lm}(\theta_1' + 2\pi n', \phi_1') Y_{lm}^*(\theta_1, \phi_1) = \frac{1}{4\pi} \sum_{k=0}^{\infty} (2k+1) P_k(\cos\gamma_n')$$

with:

$$\cos\gamma_n' = \cos\theta_1 \cos(\theta_1' + 2\pi n') + \sin\theta_1 \sin(\theta_1' + 2\pi n') \cos(\phi_1 - \phi_1')$$

and the fact that for each value of $l$, the integrals over $w$ effectively reduce to:

$$\mp \frac{i}{\pi^2} \int_0^\infty dw \frac{e^{\frac{\pi}{4M} w e^{-i\theta} \zeta_1}}{\omega^2}$$

in the ultra-violet domain $[w >> l, \infty)$, there is inherent in (31) an infinite number of quadratic divergences which occur in the angular sector for $\theta_1 = \theta_1' + 2\pi n'$ ; $\phi_1 = \phi_1'$. Since these divergences occur away from the coincidence space-time limit $(r_1 \neq r_1')$, they necessarily correspond to the infinitely many null geodesics which connect the fixed space-time point $x_1' \equiv (t_1' = 0, r_1', \theta_1', \phi_1') \equiv (0, \vec{r}_1)$ in the exterior region to the fixed space-like $r_1, \theta_1, \phi_1$-curve $C_+$ in the interior region (with $\theta_1 = \theta_1'$ ; $\phi_1 = \phi_1'$ signifying the radial null geodesic connecting $x_1'$ and $x_1$ on $C_+$). As space-like variable $t$ varies on $C_+$, it gives rise to a quadratic divergence in $D(x_1' - x_1)$ whenever space-time point $x_1$ is connected to $x_1'$ by a null geodesic. Such a behavior is in perfect conformity with the character of the massless conformal scalar propagator $D(x_1' - x_1)$ as a Hadamard function and testifies to the latter's physical consistency.

It is reiterated that as the propagator in (31) has a manifestly semi-classical character, it involves the afore-stated infinite sequence of null geodesics in the dominant $e^{-(8\pi M)w'}$-factor. The one-to-one correspondence between these infinitely many null geodesics and the afore-stated infinitely many quadratic divergences formally follows now as a direct consequence of that semi-classical character:

$$e^{-8\pi M w'} = lim_{(n\to\infty \; ; \; \theta_{\tilde{P}} \to \pi)} e^{-\frac{|\tilde{P}|}{4M}(8\pi M) n \sin\theta_{\tilde{P}}} =$$

(32) $$lim_{(n\to\infty \; ; \; \theta_{\tilde{P}} \to \pi)} e^{-\frac{|\tilde{P}|}{4M}(8\pi M)(n + \sum_{n'=0}^\infty n') \sin\theta_{\tilde{P}}}$$

This establishes that the divergence structure of the semi-classical propagator $D(x_1' - x_1)$ as an explicit function of the Schwarzschild black-hole space-time in (31) is consistent with the singularity structure of the general propagator $K(x_1, x_1')$.

## IV. HAWKING RADIATION

The entire analysis in the preceding two sections allows for the evaluation of the amplitude that a Schwarzschild black hole emit a particle which is registered in a mode of positive-definite energy $E$ by a detector at specified space-time point $x_1' \equiv (t_1' = 0, r_1', \theta_1', \phi_1')$. As already stated, the evaluation of that amplitude requires that the afore-mentioned detector be one of infinitely many which surround the black hole at some large constant radius, all of which are represented by the world-line in region $I$ of the Penrose-Carter diagram in Fig.1 [2]. In the context of the present calculation that radius lies within the range of validity of the scalar propagator $D(x_1' - x_1)$ stated in (6) which is orders of magnitude above the scale within which quantum and backreaction effects are comparatively pronounced [11]. The reason the stated range of validity is indeed orders of magnitude above the stated scale was analysed in detail in [11]. By way of further elucidating this statement, it is here reiterated that as the present treatment of Hawking radiation is predicated on the assumption in [1 - 3] to the effect that the exterior geometry of the Schwarzschild black hole remains static, the reference to the quantum and backreaction effects is made with respect to the Schwarzschild metric.

Replacing, for that matter, (31) in (18) with the inherent interpretation of $e^{-(8\pi M)w'}$ in (32), results in:

$$\mathcal{E}_E(\vec{r}_1{}', \vec{r}_1) =$$

$$e^{-8\pi M w'} \int_{-\infty}^{+\infty} dt e^{-iEt} \Big[ C e^{i\frac{p_0}{4M}n(8\pi M)} e^{\frac{p_0}{4M}t'_1} e^{i\pi p_0} |\tilde{P}| \frac{e^{-i\pi|\tilde{P}|}}{\sin(-\pi|\tilde{P}| + \pi p_0)} \frac{e^{-i\frac{|\tilde{P}|}{4M}n(8\pi M)}}{-|\tilde{P}| + p_0} e^{-i\frac{|\tilde{P}|}{4M}t'_1} \times$$

$$\frac{i}{16\pi^2 M} \frac{1}{\sqrt{\zeta_1}} \frac{1}{\sqrt{\rho'_1}} \sum_{l=0}^{\infty} \sum_{m=-l}^{l} Y_{lm}(\theta'_1 + 2\pi n', \phi'_1) Y^*_{lm}(\theta_1, \phi_1) \times$$

$$[e^{-i\frac{\pi}{4M}u_0[0]\zeta_1} \int_{\frac{\pi}{2}}^{\frac{3\pi}{2}} d\theta e^{i\theta} e^{-i\frac{3\pi}{16M}e^{-i\theta}\zeta_1} \int_0^{\infty} d\omega \frac{e^{\frac{\pi}{4M}we^{-i\theta}\zeta_1}}{\pi^2(u_0[0]e^{i\theta} + i\omega)^2 + 4(l^2 + l + 1)} -$$

(33) $\quad e^{i\frac{\pi}{4M}u_0[0]\zeta_1} \int_{\frac{\pi}{2}}^{\frac{3\pi}{2}} d\theta e^{i\theta} e^{i\frac{3\pi}{16M}e^{-i\theta}\zeta_1} \int_0^{\infty} d\omega \frac{e^{\frac{\pi}{4M}we^{-i\theta}\zeta_1}}{\pi^2(u_0[0]e^{i\theta} - i\omega)^2 + 4(l^2 + l + 1)} \Big]\Big]$

which resembles in form the general expression in (17). In order to establish that (33) is indeed the concrete expression of the latter, it is necessary to note that consequent upon the fact that $w'$ is an arbitrary real positive constant, the propagator in (33) is equivalent to:

(34) $$D(x'_1 - x_1) = e^{-(8\pi M)w''}\,[\ ]\ ;\ w'' > 0$$

with $[\ ]$ being the extended mathematical expression which constitutes the integrand up to the multiplicative factor $e^{-iEt}$ in (33). The arbitrary real positive constant $w''$ has the same physical significance as that of $w'$. In what follows, $[\ ]$ will be identified as the propagator $D(x'_1 - x_1)$ with the multiplicative factor $e^{-8\pi M w''}$ considered implicitly present. In fact, the significance of this operation extends beyond $D(x'_1 - x_1)$ as, in view of the absence of $t$-dependence in (34), the latter may also be identified as the concrete expression of the general propagator $K(t - i4\pi M, \vec{r}_1; 0, \vec{r}_1{}')$.

The issue now is the determination of $w'$ in (33). This real positive, but otherwise arbitrary, constant clearly signifies the energy of the emitted radiation registered at $x'_1$, a fact directly deduced through simple dimensional analysis. More rigorously, it follows from the fact that - pursuant to (10) - $(8\pi M)w'$ is the Euclidean semi-classical action for a single excitation of the massless conformal scalar field $\Phi(x)$. The precise relation of $w'$ to the positive-definite energy $E$ of the mode in which such an excitation is registered is just as clearly:

(35) $$w' = \frac{E}{2}$$

since the probability for particle emission by the black hole is the modulus squared of the amplitude $\mathcal{E}_E(\vec{r}_1{'}, \vec{r}_1)$ in (33).

Consequently, the amplitude in (33) is equivalent to:

$$(36)\quad \mathcal{E}_E(\vec{r}_1{'}, \vec{r}_1) = e^{-(8\pi M)\frac{E}{2}} \int_{-\infty}^{+\infty} dt e^{-iEt} D(x_1' - x_1) \equiv e^{-(8\pi M)\frac{E}{2}} \int_{-\infty}^{+\infty} dt e^{-iEt} D(x_1' - x_1'')$$

which is identical to (17) with $D(x_1' - x_1)$ now interpreted as the concrete expression $D(x_1' - x_1'')$ of the general propagator $K(t - i4\pi M, \vec{r}_1; 0, \vec{r}_1{'})$ since, again, propagation across the event horizon renders the massless conformal scalar propagator $\xi$-independent and, thereby, $t$-independent.

Thus, the entire analysis has eventuated in the central result stated in [2]: The amplitude $\mathcal{E}_E(\vec{r}_1{'}, \vec{r}_1)$ a mode of positive-definite energy $E$ propagate to fixed space-time point $x_1' \equiv (0, \vec{r}_1{'})$ in region $I$ from surface $C_+$ in region $II$ is equal to $e^{-(8\pi M)\frac{E}{2}}$ times the amplitude a mode of the same energy propagate to $x_1'$ from surface $C_-$ in region $III$ which is $C_+$-reflected in the origin of the $U - V$ plane. However, in this independent approach to Hawking radiation this result emerges directly as the exclusive consequence of the Schwarzschild black hole's event horizon and, thereby, as the exclusive consequence of the causal and topological structure of the Schwarzschild black-hole space-time. The Euclidean propagator in (12) was analytically continued back to real time in (26) and, thereby, in (31). $D(x_1' - x_1)$ in (31) is, for that matter, the concrete expression of the general propagator $K(x_1, x_1')$ whose analytical extention in the complex $t$-plane results in Hawking radiation and its thermal spectrum in [2]. In the present approach which is predicated on $D(x_1' - x_1)$ as an explicit function of the Schwarzschild black-hole space-time, no analytical extension in the complex $t$-plane is required!

This result can be best appreciated by examining the physical content of (36). The effect of infinite time-dilation on the event horizon (responsible for the centrally important $e^{-(8\pi M)\frac{E}{2}}$-factor) is the reason a distant static observer interprets the general propagator in (14) as the semi-classical propagator in (31). Since the same effect is, at once, equivalent to translating the Schwarzschild temporal coordinate $t$ by an amount $-i4\pi M$ in the complex $t$-plane and, thereby, to reflecting the Kruskal coordinates $U$ and $V$ in the origin, the event horizon of the Schwarzschild black hole is itself responsible for the intimate relation between the amplitude a mode of positive-definite energy $E$ propagate from surface $C_+$ in the interior region to point $x_1' \equiv (0, \vec{r}_1{'})$ in the exterior region of the black hole and the amplitude the same mode propagate to $x_1'$ from surface $C_-$ in the interior region of the white hole. Again, by time-reversal invariance, the modulus squared of this latter amplitude is exactly equal to the modulus squared of the amplitude the black hole absorb a particle which emanates from $x_1' \equiv (0, \vec{r}_1{'})$ in a mode of positive-definite energy $E$ and thereby eventuates at $C_+$ in a mode of the same energy. This fundamental connection between emission and absorption reveals that a black hole emits particles with a thermal spectrum characterized by temperature:

$$T_H = \frac{1}{8\pi M}$$

Hartle and Hawking generalized this intimate relation between emission and absorption to fields of spin greater than zero in the context of a reasonable simplifying assumption [2]. In the context of the present analysis, the only assumption required for the generalization of this relation to fields of spin greater than zero is that the relevant propagator feature the centrally important factor $e^{-(8\pi M)\frac{E}{2}}$. Again, the reasonable character of this assumption follows immediately from the infinite time-dilation $t \to \infty$ which the distant observer registers in the vicinity of the event horizon. This relation between emission and absorption as the exclusive consequence of the event horizon likewise generalizes to the Kerr black hole and to the Reissner-Nordström black hole [2].

Thus, from the perspective of the distant observer in the exterior region of a black hole, the event horizon is the underlying reason that [2]:

(Probability for a Schwarzschild black hole to emit a particle in a mode with energy $E$) = $e^{-(8\pi M)E} \times$ (Probability for a Schwarzschild black hole to absorb a particle in the same mode).

The preceding analysis has:

(i) Shown that the divergence structure of the semi-classical propagator $D(x'_1 - x_1)$ as an explicit function of the Schwarzschild black-hole space-time in (31) is consistent with the singularity structure of the general propagator $K(x_1, x'_1)$.

(ii) Eventuated in the independent derivation of Hawking radiation from the perspective of the semi-classical propagator in (31).

(iii) Established that from the perspective of the distant observer expressed by the semi-classical propagator in (31), Hawking radiation is the exclusive consequence of the infinite time-dilation on the event horizon.

(iv) On the basis of (i), (ii) and (iii), demonstrated the mathematical consistency and physical rigour of the semi-classical propagator in (31) and, thereby, of the propagators respectively established in the extrerior geometry in [11] and in the interior geometry in [12].

It is now desirable to demonstrate the mathematical consistency and physical rigour of the semi-classical propagator in (31) in its capacity to independently derive the correction to the thermal character of Hawking radiation established in [8] and [10].

## V. THE DETRACTION

As stated in the Introduction, the thermal spectrum of Hawking radiation is predicated on the condition of a fixed space-time geometry for $r > 2M$. The valid approximation which this condition signifies is the immediate consequence of the fact that for a Schwarzschild black hole much bigger than the Planck mass the rate of change of the metric $M^{-3}$ is very slow in comparison to the typical frequency of the radiation $M^{-1}$ [3]. The detraction from the thermal spectrum established in (33) may now be realised by relaxing this condition. The relevant derivation advances on the same concept as that which eventuated in (33).

Since as a direct consequence of infinite time-dilation on the event horizon $w'$ in (33) is an arbitrary real positive constant, it follows that $e^{-8\pi M w'}$ can be interpreted as:

$$(37) \qquad e^{-8\pi M(\frac{M-c}{M})w'} = e^{-8\pi M w' + 8\pi c w'} = e^{-8\pi (M-c)w'} \; ; \; 0 < c < M$$

on the understanding that as this operation modifies the mass of the black hole from $M$ to $M - c$, it is not permissible in the context of the condition for a fixed geometry for $r \geq 2M$. It was, in fact, shown in [8] (and elaborated in [9]) that if this condition is relaxed and energy-conservation properly taken into consideration then in the event of emission the event horizon contracts from its original $r_S = 2M$ radius to a new smaller radius $r'_S = 2(M - E)$, with $E$ being the energy of the emitted particle. The contraction of the event horizon is realised through two channels. The outgoing particle emerges "just above" the new location of the event horizon $(2(M - E) + \epsilon)$ through tunneling from "just below" the latter's original location $(2M - \epsilon)$. Equivalently, the tunneling of either particle in a pair to a point "just below" the new location of the event horizon $(2(M - E) - \epsilon)$ from a point "just above" the latter's initial location $(2M + \epsilon)$ allows the other particle to transit to spatial infinity. Remarkably, the stated amount $2E$ of horizon-contraction determines the classically inaccessible region in either tunneling channel. The distinction between the two tunneling channels has no operational significance for the distant observer in that his registration of a particle cannot determine through which of the two channels that particle was emitted. Both such channels yield an equal contribution to the tunneling probability. For an emitted s-wave at the semiclassical (WKB) limit which the registration of local fiducial observers situated close to the event horizon signifies, that is:

$$(38) \qquad \Gamma \sim e^{-8\pi E(M - \frac{E}{2})}$$

The implication inherent in (38) and the preceding considerations is clear: Energy-conservation in the context of such dynamical geometry as tunneling signifies imposes the condition that the real positive constant $c$ be identified with $w'$ in (35) if (33) is to be physically acceptable with $e^{-8\pi M w'}$ interpreted in the sense of (37). In turn, in the context of such dynamical geometry as tunneling signifies the physical significance of (33) is the amplitude[4]:

$$\mathcal{E}_E(\vec{r}_1', \vec{r}_1) =$$

$$e^{-8\pi \frac{E}{2}(M - \frac{E}{2})} \int_{-\infty}^{+\infty} dt e^{-iEt} \Big[ C e^{i\frac{p_0}{4M} n(8\pi M)} e^{\frac{p_0}{4M} t_1'} e^{i\pi p_0} |\tilde{P}| \frac{e^{-i\pi |\tilde{P}|}}{sin(-\pi |\tilde{P}| + \pi p_0)} \frac{e^{-i\frac{|\tilde{P}|}{4M} n(8\pi M)}}{-|\tilde{P}| + p_0} e^{-i\frac{|\tilde{P}|}{4M} t_1'} \times$$

---

[4] As in Planck units $M$ has dimensions of length, so does $E$ in $(M - \frac{E}{2})$. For that matter, the factor $\frac{E}{2}$ multiplying $(M - \frac{E}{2})$ has dimensions of mass [8].

$$\frac{i}{16\pi^2 M} \frac{1}{\sqrt{\zeta_1}} \frac{1}{\sqrt{\rho'_1}} \sum_{l=0}^{\infty} \sum_{m=-l}^{l} Y_{lm}(\theta'_1 + 2\pi n', \phi'_1) Y^*_{lm}(\theta_1, \phi_1) \times$$

$$[e^{-i\frac{\pi}{4M}u_0[0]\zeta_1} \int_{\frac{\pi}{2}}^{\frac{3\pi}{2}} d\theta e^{i\theta} e^{-i\frac{3\pi}{16M}e^{-i\theta}\zeta_1} \int_0^{\infty} d\omega \frac{e^{\frac{\pi}{4M}we^{-i\theta}\zeta_1}}{\pi^2(u_0[0]e^{i\theta} + i\omega)^2 + 4(l^2 + l + 1)} -$$

(39) $$e^{i\frac{\pi}{4M}u_0[0]\zeta_1} \int_{\frac{\pi}{2}}^{\frac{3\pi}{2}} d\theta e^{i\theta} e^{i\frac{3\pi}{16M}e^{-i\theta}\zeta_1} \int_0^{\infty} d\omega \frac{e^{\frac{\pi}{4M}we^{-i\theta}\zeta_1}}{\pi^2(u_0[0]e^{i\theta} - i\omega)^2 + 4(l^2 + l + 1)}]]$$

a mode of positive-definite energy $E$ propagate from surface $C_+$ in interior region $II$ to fixed space-time point $x'_1$ in exterior region $I$ through tunneling across the contracting event horizon. This amplitude:

(i) is centrally based on the semi-classical propagator in (31) which corresponds to the registration of the distant observer,

(ii) includes the above-stated equal contribution of both tunneling channels and

(iii) clearly reduces to the thermal Hawking radiation in (33) if the correction $\frac{E}{2M}$ to $e^{-(8\pi M)\frac{E}{2}}$ is neglected on the grounds that $E << M$.

It is stressed that although energy-conservation is central to the amplitude in (39), that amplitude signifies an independent equivalent derivation of Hawking radiation in the context of an evolving event horizon since its semi-classical character is not the standpoint of the local fiducial observers situated close to the event horizon in decreasing order of the Schwarzschild radial coordinate (WKB approximation in [8]), but that of infinite time-dilation registered by the distant observer.

Thus, in the realistic context of a dynamical space-time geometry, (39) signifies the fact that from the perspective of the distant observer not only is the effect of infinite time-dilation on the event horizon the underlying reason for the dominant thermal character of Hawking radiation, it is also the underlying reason for the correction $\frac{E}{2M}$ which points to an intriguing resolution of the information paradox.

## VI. CONCLUSION

Semi-classicality in the context of Hawking radiation emerges in two distinct and physically equivalent perspectives. One of the two is the perspective of the distant static observer from which the typical wavelength of Hawking radiation in the vicinity of the event horizon is of the order of the size of the event horizon itself and semi-classicality emerges as a consequence of infinite time-dilation. The other is the perspective of local observers in decreasing order of the Schwarzschild radial coordinate from which the outgoing wave reduces in the vicinity of the event horizon to an effectively point-particle of very high energy. Ideally suited for the description of the first perspective on account of its explicit reference to infinite time-dilation on the event horizon is the propagator which has been established as an explicit function of the Schwarzschild black-hole space-time. Exploiting this advantage, the analysis herein established that propagator as the concrete expression which the general Hadamard-type propagator receives in the specific frame of the distant static observer and thereby revealed that the event horizon is the underlying reason for Hawking radiation of both, a thermal spectrum and a spectrum which detracts from thermality by an amount which energy-conservation imposes. These results are consistent with Hawking radiation as the exclusive consequence of the causal structure of the Schwarzschild black-hole space-time.